\documentclass[11pt]{article} 
\usepackage{graphicx}
\usepackage{amssymb}
\usepackage{amsmath}
\usepackage{epstopdf}
\newcommand{\be}{\begin{equation}}
\newcommand{\ee}{\end{equation}}
\newcommand{\ba}{\begin{eqnarray}}
\newcommand{\ea}{\end{eqnarray}}
\begin{document}

\begin{flushright}
UCB-PTH-11/04
\end{flushright}
\vskip 1cm
\begin{center} {\Large{\textsc{Causal three-point functions} }}
\end{center}
\begin{center} {\Large{\textsc{ And} }}
\end{center}
\begin{center} {\Large{\textsc{nonlinear second-order hydrodynamic coefficients} }}
\end{center}
\begin{center} {\Large{\textsc{ in} }}
\end{center}
\begin{center} {\Large{\textsc{ AdS/CFT} }}
\end{center}
\vskip 1cm
\centerline{\bf Omid Saremi$^a$\footnote{omid.saremi@berkeley.edu} 
and Kiyoumars A. Sohrabi$^b$\footnote{kiyoumars@physics.mcgill.ca}}
\vskip 1cm
\centerline{ ${}^a $\small Berkeley Center for Theoretical Physics and Department of Physics, }
\centerline{\small University of California, Berkeley, CA 94720}
\centerline{\small and}
\centerline{\small Theoretical physics group, Lawrence Berkeley National Laboratory, Berkeley, CA 94720}
\bigskip
\centerline{${}^b$\small Physics Department, McGill University, Montr\'{e}al, QC, Canada }
\bigskip
\begin{abstract}
In the context of $\mathcal{N}=4$ SYM, we compute the finite 't Hooft coupling $\lambda$ correction to the non-linear second-order hydrodynamic coefficient $\lambda_3$ from a Kubo formula based on fully retarded three-point functions using AdS/CFT. Although $\lambda_3$ is known to vanish in the infinite 't Hooft coupling limit, we find that the finite $\lambda$ correction is non-zero. We also present a set of Kubo formulae for the non-linear coefficients $\lambda_{1,2,3}$, which is more convenient than the one that has appeared recently elsewhere. 
\end{abstract}
 \vskip 1 cm 
\section{Introduction}
The growing body of work on connections between gravity and hydrodynamics in the context of AdS/CFT seems to point towards an emerging picture: dynamics of small fluctuations of certain black-brane backgrounds is captured by the relativistic hydrodynamic equations \cite{Shiraz}. On the other hand,  AdS/CFT technology has been employed in studying strongly correlated phenomena, in particular understanding the strongly coupled quark-gluon plasma (QGP) created at RHIC. It turns out that hydrodynamics with negligible shear viscosity to entropy density ratio accurately captures the essential features of the state. Hydrodynamics is a classical effective field theory organized as a gradient expansion. When spatial gradients in units of the mean-free-path are large, higher terms in the gradient expansion, beyond first-order become increasingly important, for instance in early-time evolution of the QGP. 
On general grounds, the stress tensor of a fluid can be decomposed as 
\be\label{T}
\langle T^{\mu\nu} \rangle =  T^{\mu\nu}_{\rm eq}
 +\Pi^{\mu\nu} \,, 
 \ee 
where $T^{\mu\nu}_{\rm eq}$ is the local equilibrium piece. It is a function of the fluid four-velocity $u^{\mu}$ and energy density $\epsilon$ which itself is constrained by the equation of state. The extra piece $\Pi_{\mu\nu}$ is the dissipative part. In the present work it is truncated at the second order in derivative expansion. We work in the Landau-Liftshitz frame $u_{\mu}{\Pi^{\mu\nu}=0}$, where the separation (\ref{T}) is meaningful. The fluid velocity $u^{\mu}$ is normalized as $u_{\mu}u^{\mu}=-1$.

At second order, there is a plethora of possible operators which are
consistent with symmetries \cite{IW, Romatsch}. In a conformal field theory however, which will be our focus here, the number of independent couplings reduces down to five transport coefficients: $\tau_{\Pi}$, $\kappa$,
$\lambda_{1}$, $\lambda_{2}$ and $\lambda_{3}$, in addition to the shear
viscosity $\eta$ which already appears at first-order in gradient expansion. The ordinary effective field theory logic suggests the following form for $\Pi^{\mu\nu}$ on a general curved background \cite{Baier} 
\begin{eqnarray}
  \label{viscosity}
\Pi^{\mu\nu} & = & - \eta\sigma^{\mu\nu}
   + \eta \tau_{\Pi}\left(u\cdot \nabla
                  \sigma^{\langle \mu\nu \rangle}
                + \frac{\nabla\cdot u}{3}\sigma^{\mu\nu}\right)
 + \kappa\left( R^{\langle\mu\nu\rangle}
     - 2u_{\alpha}u_{\beta}R^{\alpha\langle\mu\nu\rangle\beta}\right)
\\ &&
+ \lambda_{1}\sigma_{\lambda}{}^{\langle\mu}
\sigma^{\nu\rangle\lambda}
+\lambda_{2}\sigma_{\lambda}{}^{\langle\mu}\Omega^{\nu\rangle\lambda}
-\lambda_{3}\Omega_{\lambda}{}^{\langle\mu}\Omega^{\nu\rangle\lambda}
  \,, \nonumber
\end{eqnarray}
where the angular brackets are defined as  $A^{\langle \mu \nu \rangle} \equiv \frac{1}{2} \Delta^{\mu\alpha}
\Delta^{\nu\beta} ( A_{\alpha\beta} + A_{\beta\alpha} ) -
\frac{1}{3} \Delta^{\mu\nu} \Delta^{\alpha\beta} A_{\alpha\beta} \,
$
and the shear tensor $\sigma^{\mu\nu} \equiv  2
\nabla^{\langle \mu } u^{\nu \rangle}$ and vorticity
$\Omega^{\mu\nu}  \equiv  \frac{1}{2} \Delta^{\mu\alpha}
\Delta^{\nu\beta} ( \nabla_\alpha u_\beta - \nabla_\beta u_\alpha)$
are written correspondingly using the angular brackets. Note that the coefficients $\lambda_{1,2,3}$ are non-linear and necessary in order for the second order hydrodynamics to be consistent \cite{Baier}. 

There have been intensive efforts in evaluating these transport coefficients in various contexts, ranging from strongly coupled $\mathcal{N}=4$ super Yang-Mills theory (SYM) with infinitely many colors and infinite 't Hooft coupling $\lambda$ \cite{Shiraz} to perturbative QCD \cite{Guy}. Interestingly, for $\mathcal{N}$=4 SYM at infinite 't Hooft coupling, the coefficient $\lambda_3$ vanishes \cite{Shiraz}. For a generic theory at arbitrary coupling, $\lambda_3$ is non-vanishing. See for example \cite{MS} for evaluation of $\lambda_3$ at  zero coupling in scalar theory.  

One convenient method to derive transport coefficients from a microscopic description is to use the so called Kubo formulae based on hydrodynamics. Recently, a set of Kubo formulae for the second order transport coefficients appeared in \cite{MS}, which extends the previously known Kubo formulae \footnote{Kubo formulae for $\tau_{\Pi}$ and $\kappa$, involving stress tensor {\it retarded two-point functions} were already known from the work of Baier $et\  al$ \cite{Baier}.} to further include the non-linear coefficients $\lambda_{1,2,3}$. These Kubo formulae are expressed in terms of {\it fully retarded three-point functions} of the stress tensor. 
In this work, first  we present our own set of Kubo formulae for the second order hydrodynamics. They also involve stress tensor three-point functions, however compared to \cite{MS} a different set of the stress tensor polarizations are chosen. Our choice proves more advantageous in a holographic setting.
As previously mentioned, $\lambda_3$ vanishes for $\mathcal{N}=4$ at $\lambda=\infty$. In this paper, using the Kubo formula we propose for $\lambda_3$, we work out the next-to-leading contribution to $\lambda_3$ at finite 't Hooft coupling. The coefficient $\lambda_3$ turns out to be {\it non-zero}. This new result is one of the central motivations behind writing this note. 
As exercises along the way, we calculate fully retarded three-point functions in AdS/CFT using the gravitational dual description of $\mathcal{N}$=4 SYM theory at infinite as well as finite coupling. More recently, proposal for computing three-point correlation functions appeared in \cite{Arnold}, which seems to be equivalent to the prescription \cite{van}. The results we find can be checked against the existing results in AdS/CFT using other methods \cite{Shiraz, Son}. This is a check on the Kubo formulae we propose (\ref{L3_k}), (\ref{k_eta}) and (\ref{L2}). This also checks the prescription of \cite{Arnold} for computing causal retarded $n$-point functions in AdS/CFT. 

The organization of the paper is as follows. In section two we define the fully retarded three-point functions. Section three is devoted to explaining the setup. The Bulk three-point vertices and computation of the back-reactions on the metric, necessary for calculating the three-point functions, are discussed in section four. Then we move on to the finite 't Hooft coupling corrections and compute $\lambda_3$ at next-to-leading order. We conclude the presentation with a discussion. 
\section{Causal 3-point functions }
We will only be working with $\mathcal{N}$=4 SYM in four-dimensions (which implies a five-dimensional bulk) to avoid lengthly dimension-dependent expressions among other reasons. In what follows, Greek letters denote the boundary spacetime coordinates. These coordinates will also be referred to as $(t,x,y,z)$ interchangeably. Uppercase Latin letters $M,N,\cdots$ are reserved for the five-dimensional bulk spacetime. 

There are various ways of defining fully retarded three-point functions; one is through functional derivatives of the real-time generating functional $W$ with respect to the source terms. Alternatively, one could define them without introducing sources and merely by performing a path integral. In general, the two definitions above will differ by contact terms \cite{Son}. The definition which is more intimately connected to the
gauge/gravity correspondence is indeed to use the effective
action. In \cite{Baier, MS}, the closed-time-path formalism was employed to define the generating functional $W$ 
\begin{eqnarray} \!\!\!\!\!
e^{W[h_1,h_2]} & \equiv &  \int {\cal D}[\Phi_1,\Phi_2,\Phi_3]
e^{i\int \sqrt{-g_1} d^4 x {\cal L}[\Phi_1(x),h_1]} \nonumber \\ &&
\times e^{-\int_0^\beta \!\!\! d^4 z {\cal L}_{\rm E} [\Phi_3(z)]}
e^{-i\int \sqrt{-g_2} d^4 y {\cal L}[\Phi_2(y),h_2]} \ , \nonumber
\end{eqnarray}
where $h_{1}$ and $h_{2}$ are the metric fluctuations (i.e., source terms for the stress tensor) living on the upper and lower closed-time contour respectively. See \cite{MS} for more details on definitions. It is common to choose a new field basis (the so called $ra$-basis) by defining average sources and operators 
$h_{r}\equiv\frac{h_{1}+h_{2}}{2}$ and 
$T_{r}\equiv\frac{T_{1}+T_{2}}{2}$ as well as difference variables
$h_{a}\equiv h_{1}-h_{2}$ and $T_{a}\equiv T_{1}-T_{2}$. Variation of the effective action with respect to $h_{a}$
yields $T_{r}$
\begin{eqnarray}\label{Res}
\langle T^{\mu\nu}_r(x) \rangle= 
\frac{-2i}{\sqrt{-g}} \frac{\partial W}{\partial h_{a\mu\nu}(x)},
\end{eqnarray}
where one sets $h_{a}=0$ and
$h_{r}=h_{1}=h_{2}=h$ (the classical background value) only after performing the functional derivative. More generally, the fully retarded $n$-point function is defined as 
\begin{eqnarray} \hspace{-0.2em}
G^{\mu\nu,\alpha\beta,\ldots}_{r a \ldots}(0,x,\ldots) &
\!\equiv\!\! & \left. \frac{(-i)^{n-1}(-2i)^n \partial^n W}{\partial
g_{a,\mu\nu}(0)
         \partial g_{r,\alpha\beta}(x) \ldots }
        \right|_{g_{\mu\nu}=\eta_{\mu\nu}}
\\ & \!=\!\! & (-i)^{n-1}
\left\langle T^{\mu\nu}_r(0) T^{\alpha\beta}_a(x) \ldots
     \right\rangle_{\rm eq} \!
+\mbox{c.t.},  \nonumber
\end{eqnarray}
where c.t. stands for possible contact terms. Before proceeding, let us spell out our convention for Fourier transformation
\be\label{fdef}
G^{\mu\nu,\alpha\beta,\rho\lambda}_{raa}(0,y,z)=\int d^4p d^4q~G^{\mu\nu,\alpha\beta,\rho\lambda}_{raa}(p,q)e^{ip\cdot y+iq\cdot z },
\ee
where $p\cdot x=-p_0 t+p_z z$. The calligraphic letters say $\mathfrak{p}$ are defined as $\mathfrak{p}=p/(2\pi T)$, where $p$ is a momentum and $T$ is the temperature of the system.

We are now in a position to present the new set of Kubo formulae, involving stress-tensor three-point functions for the second-order hydrodynamic coefficients. 
Consider hydrodynamics in flat space in local equilibrium. By perturbing the system through small background metric disturbances $g_{\mu\nu}=\eta_{\mu\nu}+h_{\mu\nu}$, one can create shear
flow $\sigma^{\mu\nu}$ and/or vorticity $\Omega^{\mu\nu}$. Solving the hydrodynamics equations of motion, namely conservation law for the stress tensor, one can obtain a consistent fluid velocity  background $u^{\mu}$ as a function of $h_{\mu\nu}$ and determine the expansion of $\langle
T^{\mu\nu}_{r}\rangle$ in powers of $h_{\mu\nu}$ from the constitutive relations. Field theoretically on the other hand, one can expand the response of $\langle T^{\mu\nu}_r\rangle$  to the perturbations in terms of the fully retarded $n$-point correlation functions 
\ba\label{def}
\langle T^{\mu\nu}_r(x)\rangle&=&G^{\mu\nu}_{ra}(x)-\frac{1}{2}\int d^4y G^{\mu\nu, \alpha\beta}_{ra}(x,y)h_{\alpha\beta}(y) \\\nonumber &&+\frac{1}{8}\int d^4y d^4z G^{\mu\nu,\alpha\beta,\rho\lambda}_{raa}(x,y,z)h_{\alpha\beta}(y)h_{\rho\lambda}(z)+\mathcal{O}(h^3).
\ea
The following set of (flat boundary) metric fluctuations can be consistently turned on and results in convenient Kubo formulae: $\{h_{xy}(t), h_{xz}(t), h_{yz}(t), h_{tx}(z),h_{ty}(z)\}$. Let us work out $\lambda_3$ as an example 
\begin{eqnarray}
\langle T^{xy}_r\rangle=-\frac{\kappa}{2}h_{ty}\partial^2_{z}h_{tx}-\frac{\kappa}{2}h_{tx}\partial^2_{z}h_{ty}+\frac{\lambda_{3}}{4}\partial_{z}h_{tx}\partial_{z}h_{ty}+\mathcal{O}{(h^3)},\quad u^{\mu}=(1,0,0,0)+\mathcal{O}(h),
\end{eqnarray}
where only contributions from $h_{tx/ty}$ are kept. Recalling the definition (\ref{def}), one obtains 
\begin{eqnarray}\label{L3_k}
\lambda_{3}=-4\lim_{p,q\rightarrow0}\partial_{pz}\partial_{qz}G^{xy,tx,ty}_{raa}(p,q), \quad \kappa=\lim_{p,q\rightarrow 0}\frac{\partial^2}{\partial p_z^2}G_{raa}^{xy, tx,ty}(p,q).\quad 
\end{eqnarray}
Similarly, one can extract the following relations
\be\label{k_eta}
\eta=i\lim_{p,q\rightarrow 0}\frac{\partial}{\partial q_0}G^{xy,xz,yz}_{raa}(p,q),\quad 2\eta\tau_{\pi}-\kappa=\lim_{p,q\rightarrow 0}\frac{\partial^2}{\partial p_{0}^2} G_{raa}^{xy,xz,yz}(p,q), 
\ee
\begin{equation*}
\quad \lambda_{1}=\eta\tau_{\pi}-\lim_{p,q\rightarrow 0}\frac{\partial^2}{\partial p_{0}\partial q_{0}} G_{raa}^{xy, xz,yz}(p,q),
\end{equation*}
where $p_0+q_0\neq 0$ in the Kubo formula for $\eta$. We also have
\be\label{L2}
\quad 
\lambda_{2}=2\eta\tau_{\pi}-4\lim_{p,q\rightarrow 0}\frac{\partial^2}{\partial p_{0}\partial q_{z}} G_{raa}^{xy, ty,xz}(p,q).
\ee
\section{The setup}
In this section we will be concerned with the stress-tensor correlation functions in the infinite 't Hooft coupling limit, thus it is sufficient to restrict ourselves to the two-derivative sector of IIB supergravity action involving only gravity and a negative cosmological constant.\footnote{We will not be considering loops. The action (\ref{action}) will contain the full set of relevant bulk vertices.} Later in the text, we manage to compute next-to-leading finite 't Hooft coupling contribution to $\lambda_3$. As mentioned before, this is interesting since the leading contribution is zero. To accomplish this task, one is required to go beyond the two-derivative action (\ref{action}). More explanations will appear in the section (\ref{finite}). For the present section the action, including counter-terms is 
\be\label{action}
\mathcal{S}=\frac{N^2}{8 \pi^2 L^3}[\int d^5x\sqrt{-g} (R+\frac{12}{L^2})+2\int d^4x \sqrt{-h}~K-\frac{6}{L}\int d^4x \sqrt{-h}].
\ee
The background solution representing thermal equilibrium in the boundary theory is a Schwarzschild-AdS black hole with a planar horizon. The following coordinate system is convenient   
\be
ds^2=g^{(0)}_{MN}dx^{M}dx^{N}=\frac{(\pi T L)^2}{u}[-fdt^2+dx^2+dy^2+dz^2]+\frac{L^2}{4u^2f}du^2,
\ee
where Hawking temperature of the hole is $T$, $L$ is the AdS radius and $f=1-u^2$. The black hole horizon is at $u=1$ and the boundary of AdS sits at $u=0$. The stress tensor in asymptotically AdS$_5$ space is given by
\be\label{st}
T^{\mu\nu}=-\frac{N^2}{4\pi^2 L^3}(K^{\mu\nu}-h^{\mu\nu}K)-\frac{3N^2}{4\pi^2L^4}(h^{\mu\nu}-\frac{L^2}{6}G^{\mu\nu}),
\ee
where $\langle T^{\mu\nu}\rangle=\lim_{u\rightarrow 0}\frac{(\pi T L)^2}{u}\sqrt{-h}T^{\mu\nu}$ and $h_{\mu\nu}$ is the induced metric on the boundary at $u=0$. The extrinsic curvature of the boundary is defined as $ K_{\mu\nu}=-\frac{1}{2}(\nabla_{\mu}n_{\nu}+\nabla_{\nu}n_{\mu})$, where $n^{\mu}$ is the unit vector normal to the boundary pointing outward and $K$ is the trace of $K_{\mu\nu}$. 
The Einstein tensor $G_{\mu\nu}=R_{\mu\nu}-\frac{1}{2}g_{\mu\nu}R$ is built out of the induced metric on the boundary $h_{\mu\nu}$. 
\subsection{Recipe for three-point functions}
From now on we set the AdS radius $L=1$. There are multiple but equivalent ways to write down the bulk cubic vertices (or equivalently the full three point amplitude). One approach is to work out the on-shell action for the cubic metric fluctuations. This is the route taken for instance in \cite{Frolov}. In this approach, one has to pay great attention to the boundary terms generated in the process. However, there exists an entirely equivalent approach. Although completely straightforward, the next few lines will outline this. The functional derivative $\mathcal{V}_3=\frac{\delta^3 \mathcal{S}_{bulk}}{\delta H^{(b)}\delta H^{(b)} \delta H^{(b)}}$ evaluated on-shell directly yields the three-point correlation function, where we have suppressed the indices, $H^{(b)}$ is the boundary data and $\lim_{u\rightarrow 0}\delta g_{\rho\lambda}=\frac{(\pi T)^2}{u}\delta H^{(b)}_{\rho\lambda}$. Varying the action once leads to 
\ba\label{V3}
\delta \mathcal{S}_{bulk}=\int d^5x \sqrt{-g}\mathcal{E}_{MN}[\mathcal{G}_{PQ}]\delta g^{MN} +\int d^4x{\sqrt{-h}T^{\mu\nu}_{ren}\delta g_{\mu\nu}}_{|u=0}&=&\\\nonumber
\int d^5x \sqrt{-g}\mathcal{E}_{MN}[\mathcal{G}_{PQ}]\delta g^{MN} +\langle T^{\mu\nu}\rangle[\mathcal{G}_{\rho\lambda}]\delta H^{(b)}_{\mu\nu},
\ea     
where $\mathcal{E}_{MN}$ is the equation of motion and $\mathcal{G}_{MN}$ is an arbitrary background. Observe that if the equations of motion for the metric perturbations are solved up to second-order in fluctuations, namely, $\mathcal{E}_{MN}[g^{(0)}, \delta g]=\mathcal{O}{( \delta g^{3})}$, the first term in (\ref{V3}) will not contribute to $\mathcal{V}_3$ and only two functional derivatives of the last term will survive. Since the last term is already a boundary term, no bulk integration over the radial direction ever shows up. This will be the definition for the three-point functions, we work with in this paper. As for the external wavefuctions, the usual boundary-to-bulk propagators  (and their complex conjugates) are all that is required. We work them out in the next section. As emphasized in \cite{Omid}, the full Kruskal-Szekeres extension of the black-brane background (to introduce Schwinger-Keldysh doubler fields) is irrelevant. The fully retarded functions we are concerned with here are causal quantities, therefore, on physical grounds only one causally-connected quadrant of the fully extended AdS-Schwarzschild spacetime should suffice for computing them.  
\subsection{Solving the equations of motion with metric back-reaction}
As alluded to previously, in order to compute the stress-tensor three-point functions, it is required to evaluate the second-order back-reaction on the linearized metric fluctuations. This is formulated as 
\be\label{pert}
g_{\mu\nu}=g^{(0)}_{MN}+h_{MN}=g^{(0)}_{MN}+\epsilon h^{(1)}_{MN}+\epsilon^2 h^{(2)}_{MN},~~
\mathcal{E}_{MN}[h^{(1)},g^{(0)}]=0,\quad  \mathcal{E}_{MN}[h^{(2)}, g^{(0)}]= S[(h^{(1)})^2],
\ee
where $\epsilon$ is a small parameter indicating the order in perturbation theory and $\mathcal{E}_{MN}[h^{(1)},g^{(0)}]$ is the Einstein' s equation for the linear gravitons on the black-brane background. The term $S[(h^{(1)})^2]$ is the non-linear source stemming from interactions of the linear metric perturbations. Both $h^{(1)}$ and $h^{(2)}$ are taken to be incoming at the horizon (if $h_{\mu\nu}$ is the largest-time insertion in the three-point function).  At the boundary $h^{(1)}=H^{(b)}$ and $h^{(2)}=0$ conditions are imposed, where $H^{(b)}$ is the boundary data for $h^{(1)}$. We demand regularity for all fluctuations at the horizon, although it turns out below that for certain metric polarizations demanding regularity at the horizon must be replaced by a stronger condition dictated by the equation of motion. 
The following metric polarizations are switched on  
\be\label{metric_pert}
h_{tx}(u,z)=\frac{(\pi T L)^2}{u}H_{tx}(u)e^{ip_z z},\quad h_{ty}(u,z)=\frac{(\pi T L)^2}{u}H_{ty}(u)e^{iq_z z}.
\ee
\begin{equation*}
h_{xz}(t,u)=\frac{(\pi T L)^2}{u}H_{xz}(u)e^{-iq_0 t},\quad h_{yz}(t,u)=\frac{(\pi T L)^2}{u}H_{yz}(u)e^{-ip_0 t}.
\end{equation*}
\begin{equation*}
h_{xy}(t,z,u)=\frac{(\pi T L)^2}{u}H_{xy}(t,z,u),\quad
\end{equation*}
where the four-momentum dependence of $H_{xy}$ is selected such that momentum is conserved for the cubic vertex of under study. This set of metric fluctuations is convenient as it can be consistently extended into the bulk to satisfy the bulk equations of motion and at the same time fluctuations are decoupled from each other. Let us focus on the three-point correlator $G^{xy,yz,xz}_{raa}(p,q)$. This will be relevant to Kubo formulae (\ref{k_eta}). The ansatz for $H_{xy}(t,z,u)$ for this correlation function is 
\be
H_{xy}(t,z,u)=H_{xy}(u)e^{-i(p_0+q_0)t}.
\ee
The equation of motion for $H_{xy}(u)$ including the non-linear source term is
\be 
H^{''}_{xy}(u)+\frac{u^4-1}{uf^2}H^{'}_{xy}(u)+\frac{(\mathfrak{p}_0+\mathfrak{q}_0)^2}{uf^2}H_{xy}(u)=\frac{\mathfrak{p}_{0}\mathfrak{q}_0}{uf^2}H_{xz}(u)H_{yz}(u)+H^{'}_{xz}(u)H^{'}_{yz}(u).
\ee
It is easy to work out the equation of motion for $H_{xz}(u)$  
\be
H^{''}_{xz}(u)-\frac{1+u^2}{uf}H^{'}_{xz}(u)+\frac{\mathfrak{q}_{0}^2}{uf^2}H_{xz}(u)=0,
\ee
where $H_{yz}$ satisfies the same equation but with $\mathfrak{q}_0$ replaced by $\mathfrak{p}_{0}$. Given that we are interested in fully retarded three-point correlation functions with $xy$-polarization as the largest-time insertion, we only need to include second-order back-reaction for $H_{xy}$ (created by a non-linear source in $H_{xz/yz}$ perturbations). Back-reaction on $H_{xz/yz}$ will contribute to other $n$-point functions, which we are not of interested in the present work. 

It will be sufficient to know the wavefunction for various metric polarizations as an expansion in frequency and momentum up to second-order. Let us define
\ba
\mathcal{K}(u,\mathfrak{w})&=&(1-u)^{-i\frac{\mathfrak{w}}{2}}\large\{1-i\frac{\mathfrak{w}}{2}\ln(1+u)\\\nonumber && +\mathfrak{w}^2[\frac{1}{8}\ln^2(1+u)+(1-\frac{1}{2}\ln(2))\ln(1+u)-\frac{1}{2}{\rm dilog}(\frac{1}{2}+\frac{u}{2})+\frac{\pi^2}{24}-\frac{1}{4}\ln^2(2)]\large\},
\ea
then
\be\label{sol1}
H^{(1)}_{xy}=H^{(b)}_{xy}\mathcal{K}(u,\mathfrak{p}_0+\mathfrak{q}_0), \quad H^{(1)}_{xz}(u)=H^{(b)}_{xz}\mathcal{K}(u,\mathfrak{q}_0), \quad H^{(1)}_{yz}(u)=H^{(b)}_{yz}\mathcal{K}(u,\mathfrak{p}_0).
\ee
The set of solutions (\ref{sol1}) solves the first-order equations in (\ref{pert}). As for the back-reacted metric at the second-order, one obtains
\ba\label{sol2}
H^{(2)}_{xy}&=&-H^{(b)}_{xz}H^{(b)}_{yz}(1-u)^{-i\frac{\mathfrak{p_0}+\mathfrak{q_0}}{2}}\mathfrak{p}_{0}\mathfrak{q}_0\large[ \frac{1}{4}\ln^2(1+u)\\\nonumber &&+(1-\frac{1}{2}\ln(2))\ln(1+u)-\frac{1}{2}{\rm dilog}(\frac{1}{2}+\frac{u}{2})+\frac{\pi^2}{24}-\frac{1}{4}\ln^2(2)\large].
\ea
We need the renormalized holographic stress tensor (\ref{st}), expanded to the second-order in metric fluctuations. After substituting the solutions into the stress tensor and taking two derivatives with respect to the boundary data, one ends up with the following expression for the three-point function $G^{xy,xz,yz}_{raaa}(p,q)$
\ba\label{xyxzyz}
G_{raa}^{xy,xz,yz}(p,q)&=&\frac{\pi^2N^2 T^4}{8}-i\frac{\pi}{8}(p_0+q_0)N^2 T^3\\\nonumber &&+\frac{1}{16}(1-\ln(2))(p_0^2+q_0^2+p_0q_0)N^2T^2+\mathcal{O}(p_{0}^3, p_0q_0^2,p_0^2q_0,q_0^3).
\ea
Now let us turn our attention to $G^{xy, tx,ty}_{raa}(p_z,q_z)$. $H_{xy}(t,z,u)$ will have the following form 
\be
H_{xy}(t,z,u)=H_{xy}(u)e^{i(p_z+q_z)z}.
\ee
The equation of motion for $H_{tx/ty}$ fluctuation is 
\be\label{tx_ty}
H^{''}_{tx/ty}(u)-\frac{1}{u}H^{'}_{tx/ty}(u)+\frac{\mathfrak{Q}^2}{u(u^2-1)}H_{tx/ty}(u)=0,
\ee
where $\mathfrak{Q}$ stands for the momentum of the corresponding mode according to the definitions in (\ref{metric_pert}). Note that there is an important subtlety regarding boundary conditions here. The equation of motion (\ref{tx_ty}) dictates a stronger boundary condition at the horizon; rather than just asking for $H_{tx/ty}$ to be regular, one needs to set $H_{tx/ty}=0$ at the horizon to obtain a regular solution. We need the following wavefunctions
\be
H^{(1)}_{xy}=H^{(b)}_{xy}[1-(\mathfrak{p}_z+\mathfrak{q}_z)^2\ln{(1+u)}]+\mathcal{O}((\mathfrak{p}_z+\mathfrak{q}_z)^4),
\ee
and for $H_{tx/ty}$
\be
H^{(1)}_{tx/ty}=H^{(b)}_{tx/ty}[(1-u^2)-u(1-u)\mathfrak{Q}^2]+\mathcal{O}(\mathfrak{Q}^4),
\ee
where $\mathfrak{Q}$ stands for the spatial momentum of the mode according to (\ref{metric_pert}). The back-reaction piece reads
\ba
H^{(2)}_{xy}&=&H^{(b)}_{tx}H^{(b)}_{ty}[u^2+(\mathfrak{p}_z+\mathfrak{q}_z)^2(u-u^2-\ln{(1+u)})\\\nonumber &&+(2u^2-3u+2\ln{(1+u)})\mathfrak{p}_z\mathfrak{q}_z+\mathcal{O}(\mathfrak{p}^3_z,\mathfrak{p}_z\mathfrak{q}^2_z,\mathfrak{p}^2_z\mathfrak{q}_z,\mathfrak{q}^3_z).
\ea
evaluating the $xy$-component of the stress-tensor from (\ref{st}), we get the following expression for the fully retarded function
\be\label{xytytx}
 G^{xy,tx,ty}_{raa}(p_{z},q_{z})=\frac{N^2T^2}{16}(p^2_z+q^2_z)-\frac{\pi^2N^2T^4}{8}.
\ee
The momentum independent terms in (\ref{xytytx}) and (\ref{xyxzyz}) can be understood in the following way. The $xy$-component of the energy-momentum tensor gives
\begin{eqnarray}\label{P}
T^{xy}=(\epsilon+P)u^{x}u^{y}+Pg^{xy}+\mathcal{O}(\nabla).
\end{eqnarray}
where the error term denotes the gradient expansion. In addition, given we only care about second-order metric fluctuations, the first term in (\ref{P}) in zero frequency and spatial momentum limit will not contribute. As for the term proportional to pressure $P$, note that we
have $g^{xy}=-h_{xt}h_{yt}+\mathcal{O}(h^3)$ (and $g^{xy}=h_{xz}h_{yz}+\mathcal{O}(h^3)$), therefore
\begin{eqnarray}
\lim_{p_{z}\rightarrow0}\lim_{q_{z}\rightarrow0}
G^{xy,tx,ty}_{raa}(p,q)=-P, \quad \lim_{p_{0}\rightarrow0}\lim_{q_{0}\rightarrow0}
G^{xy,xz,yz}_{raa}(p,q)=P.
\end{eqnarray}
Similarly we show the $\lambda_2$ Kubo relation (\ref{L2}) correctly reproduces the already known results. The relevant ansatz for $H_{xy}(t,z,u)$ is
\be
H_{xy}(t,z,u)=H_{xy}(u)e^{-ip_0t+iq_z z}.
\ee
The solution for $H_{xy}$, including back-reaction from $H_{ty}$ and $H_{xz}$ is straightforward to compute
\be
H_{xy}=H^{(b)}_{xy}(1-u)^{-i\frac{\mathfrak{p}_0}{2}}[1-(\frac{i}{2}\mathfrak{p}_0+\mathfrak{q}^2_z)\ln{(1+u)}]-H^{(b)}_{xz}H^{(b)}_{ty}(1-u)^{-i\frac{\mathfrak{p}_0}{2}}\mathfrak{p}_0\mathfrak{q}_z\ln{(1+u)}.
\ee
The $xy$-component of the stress-tensor yields  
\be\label{xyxzty}
 G^{xy,xz,ty}_{raa}(p_0,q_{z})=\frac{N^2T^2}{16}p_0q_z.
\ee
Using values of the second-order hydrodynamic coefficients calculated elsewhere \cite{Shiraz, Son}
\be\label{cof}
\eta=\frac{\pi N^2T^3}{8},\quad \kappa=\frac{\eta}{\pi T}, \quad \tau_{\pi}=\frac{2-\ln{(2)}}{2\pi T},\quad  \lambda_{1}=\frac{\eta}{2\pi T},\quad  \lambda_{2}=-\frac{\eta\ln{(2)}}{\pi T},\quad \lambda_3=0,
\ee
it is now trivial to see that, the Kubo formulae  (\ref{L3_k}), (\ref{k_eta}) and (\ref{L2}), using the expressions for the three-point functions we have computed in (\ref{xyxzyz}),  (\ref{xytytx}) and (\ref{xyxzty}) reproduce the second order coefficients (\ref{cof}). This concludes the warmup exercises. 
\section{Finite 't Hooft coupling corrections}\label{finite}
In this section we turn our attention to $\lambda_3$. Our goal will be to compute next-to-leading contribution to $\lambda_3$ (finite 't Hooft coupling correction). We need to consider $\alpha^{'}$ corrected IIB supergravity action. One known correction term is at order $\mathcal{O}(\alpha^{'3})$ and is quartic in the Weyl tensor. In \cite{P1}, corrections involving F$_5$ field strength were computed and further shown not to contribute to the corrected near extremal black D3-brane solution. In another work \cite{B2}, authors show that there is no need to work with the full 10-D action, if one is only concerned with thermodynamics and/or hydrodynamic response. Therefore we use the following $\alpha^{'}$ corrected gravity action
\be\label{actionhd}
S=\frac{N^2}{8\pi^2}\int d^5x \sqrt{-g}(R+12+\gamma \mathfrak{W}), \ee
where 
\be
\mathfrak{W}=\mathfrak{B}_{\alpha\beta\rho\sigma}(2\mathfrak{B}^{\alpha\rho\sigma\beta}-\mathfrak{B}^{\sigma\alpha\beta\rho}), \quad \mathfrak{B}_{\alpha\beta\rho\sigma}=C^{\mu}_{~\alpha\beta\nu}C^{\nu}_{~\sigma\rho\mu}, 
\ee
 $C$ is the Weyl tensor and $\gamma=\frac{\alpha^{'3}}{8}\zeta(3)$. The planar AdS-Schwarzschild, in the presence of the higher derivative term is corrected. The modified background looks like
\be
ds^2=\frac{(\pi T_0)^2}{u}(-f Z_tdt^2+dx^2+dy^2+dz^2)+Z_u\frac{du^2}{4u^2(1-u^2)},
\ee
where
\be
Z_t=1-15\gamma(5u^2+5u^4-3u^6)+\mathcal{O}(\gamma^2), \quad Z_u=1+15\gamma(5u^2+5u^4-19u^6)+\mathcal{O}(\gamma^2).  
\ee
The Hawking temperature associated to the hole is now 
\be
T=T_0(1+15\gamma).
\ee

According to the rules of AdS/CFT, world-sheet corrections proportional to $\gamma$ are mapped into finite 't Hooft coupling $\lambda$ corrections using $\alpha^{' }/L^2=\lambda^{-1/2}$. Below, we find the $\gamma$ correction term to the fully retarded three-point function $G^{xy,tx,ty}_{raa}$. Using the Kubo formula (\ref{L3_k}), we find corrections to $\kappa$ as well as next-to-leading piece of $\lambda_3$. The latter turns out to be {\it non-vanishing}. The correction piece to $\kappa$ has already been calculated elsewhere in \cite{B1}. Consequently, checking our result against that of \cite{B1} is an indirect check on our $\lambda_3$ computation. 
Calculations are considerably tedious. Up to first order in $\gamma$, the solution for $H_{xy}$ is given by
\ba
H_{xy}&=&H^{(b)}_{xy}\large[1-(\mathfrak{p}_z+\mathfrak{q}_z)^2\ln{(1+u)}\\\nonumber &&+(\mathfrak{p}_z+\mathfrak{q}_z)^2\gamma(\frac{43}{2}u^6-65u^5+\frac{135}{2}u^4-\frac{250}{3}u^3+\frac{195}{2}u^2+175\ln{(1+u)}-175u)\large].
\ea
The solution for $H_{tx/ty}$ reads
\ba
H_{tx}&=&H^{(b)}_{tx}[(1-u^2)-15\gamma u^2(1-u^2)(-3u^4+5u^2+5)-\mathfrak{p}^2_zu(1-u)\\\nonumber &&-5\gamma\mathfrak{p}^2_z u^2(1-u)(9u^5+25u^4+u^3+u^2+u+1)],
\ea
and similarly for $H_{ty}$ with $\mathfrak{p}_z$ replaced by $\mathfrak{q}_z$, where boundary condition $H_{tx/ty}=0$ at the horizon was imposed.
Given these first-order solutions, one can find the second-order back-reaction on $H_{xy}$. We need to derive the $H_{xy}$ equation of motion by varying the action (\ref{actionhd}), including the non-linear quadratic source terms coming from $H_{tx/ty}$. We do not write the resulting equation of motion down, as it is extremely long and non-illuminating. We present the wavefunctions, as they may be useful for future explorers  of the topic. To spell out the solution more efficiently, let us define
\be
H^{(2)}_{xy}=H^{(b)}_{tx}H^{(b)}_{ty}\large[F_1+\gamma F_2+(\mathfrak{p}_z+\mathfrak{q}_z)^2F_3-\mathfrak{p}_z\mathfrak{q}_zF_4+\gamma(-\mathfrak{p}_z\mathfrak{q}_zG_1+(\mathfrak{p}_z+\mathfrak{q}_z)^2G_2)\large]
\ee
where 
\be
F_1=u^2,\quad F_2=15u^2(u^2-1)(3u^4-5u^2-5),\quad F_3=u-u^2-\ln{(1+u)},
\ee
\be
F_4=-2u^2+3u-2\ln{(1+u)},
\ee
\be
G_1=-90u^8-\frac{610}{7}u^7+283u^6-140u^5+135u^4-\frac{550}{3}u^3+205u^2-400u+400\ln{(1+u)},
\ee
\be
G_2=-45u^8-80u^7+\frac{283}{2}u^6-65u^5+\frac{135}{2}u^4-\frac{250}{3}u^3+\frac{205}{2}u^2-175u+175\ln{(1+u)}.
\ee
At this point we must comment on the stress tensor in the presence of the quartic Weyl tensor contribution. We find that the stress tensor (\ref{st}) is not affected by the higher derivative correction in (\ref{actionhd}). Indeed, contributions from the background proportional to $\gamma$ fall off too fast to make non-zero contribution to the stress tensor. So even in the presence of the term proportional to $\gamma$ in (\ref{actionhd}), the expression for the stress tensor remains unchanged.  A similar observation was made in \cite{B3}.

Here is the three-point function, including the finite $\lambda$ correction
\be
G^{xy,tx,ty}(p_z,q_z)=-\frac{\pi^2 N^2 T^4}{8}-\frac{15\pi^2 N^2 T^4}{8}\gamma+\frac{N^2T^2}{16}(p^2_z+q^2_z)-\frac{5N^2T^2}{8}(p^2_z+q^2_z-5p_zq_z)\gamma+\mathcal{O}(\gamma^2).
\ee 
From this, for vanishing $p_z$ and $q_z$, we recover the famous result of \cite{Gubser}, i.e.,  the finite 't Hooft coupling correction to the free-energy density of $\mathcal{N}=4$ SYM at finite temperature. Using the Kubo formula (\ref{L3_k}) we obtain 
\be
\kappa=\frac{N^2T^2}{8}-\frac{5N^2T^2}{4}\gamma,
\ee
which agrees with what appeared in \cite{B1}. Having passed these checks, we can read off $\lambda_3$ using the Kubo relation (\ref{L3_k})
\be\label{L3R}
\lambda_3=-\frac{25N^2 T^2}{2}\gamma=-(\frac{5}{4})^2\zeta(3)\lambda^{-\frac{3}{2}} N^2 T^2.
\ee
This result, along with the set of Kubo formulae (\ref{L3_k}), (\ref{k_eta}) and (\ref{L2}) are main results of this note. Note that the minus sign is just our convention used in writing the term proportional to $\lambda_3$ in (\ref{viscosity}). 
\subsection{ Ward identities}
In this section, we provide further check on our $\lambda_3$ result (\ref{L3R}). The coordinate invariance of the $\mathcal{N}$=4 generating functional implies Ward identities. One particular Ward identity presented in \cite{MS}, involves both two and three-point functions and only one momentum
\be\label{Ward}
[\eta^{\mu\gamma}G^{\delta \nu,\alpha\beta}_{ra}(Q)+(\mu\leftrightarrow\nu)]+[\eta^{\alpha\gamma}G^{\mu\nu, \delta\beta}_{ra}(Q)+(\alpha\leftrightarrow\beta)]+(\gamma\leftrightarrow\delta)=2G^{\mu\nu,\alpha\beta,\gamma\delta}_{raa}(Q,0).
\ee
Note that we could always arrange for one four-momentum to vanish in the correlator relevant to the $\lambda_3$ computation; one could pick $q_{z}=-p_{z}$ and $p_0=q_0=0$. From (\ref{Ward}) we find 
\be\label{useWard}
G^{tx,tx}_{ra}(p)=G^{tx,ty,xy}_{raa}(p,0),
 \ee 
where the fact that $G^{tx,tx}(p) =G^{ty,ty}(-p)$ in our setup was used. Note that after Euclidean continuation of (\ref{useWard}), one can reinterpret the polarizations and write \footnote{After going Euclidean both sides pick a minus sign.}
 \be\label{final}
G^{tx,tx}_{E}(p_z)=G^{xy,tx,ty}_{E}(p_z,-p_z). 
 \ee 
The correlator $G^{xy,tx,ty}(p_z,-p_z)$ obviously contains less information than the full correlator with the two momenta independent but it is easy to check (using hydrodynamics) that it is sensitive to a combination of $\kappa$ and $\lambda_3$
\be\label{expr}
G^{xy,tx,ty}_{E}(p_z,-p_z)=-P+(\kappa+\frac{1}{2}\lambda_3)p_z^2+\mathcal{O}(p_z^3).
\ee
The Ward identity (\ref{final}) indicates that $G^{tx,tx}_{E}(p_z)$ should know about a combination of $\lambda_3$ and $\kappa$ according to (\ref{expr}). We evaluate the Euclidean two-point function in (\ref{final}) and find that indeed the Ward identity (\ref{final}) holds
\be
G^{tx,tx}_{E}(p_z)=G^{xy,tx,ty}_{E}(p_z,-p_z)=-\frac{\pi^2N^2T^2}{8}+\frac{N^2T^2}{8}p^2_z-(\frac{35N^2T^2}{8}p^2_z+\frac{15\pi^2N^2T^4}{8})\gamma+\mathcal{O}(\gamma^2).
\ee
This is yet another non-trivial test on our result, equation (\ref{L3R}).
\section{Discussion}
In this section, we discuss implications of our findings. We also comment on future directions. It was known from \cite{Shiraz}, that $\lambda_3=0$ for the infinitely coupled $\mathcal{N}=4$ SYM plasma. Our results indicate that at finite but large values of $\lambda$, the coefficient $\lambda_3$ is non-vanising. In \cite{MS}, it was noted that for generic weakly coupled field theories, $\lambda_{3}$ is non-zero. Although the same computation for $\mathcal{N}=4$ has not yet been completed \cite{Sohrabi}, it is likely that the same is true about $\mathcal{N}=4$ SYM at weak or zero coupling (perturbative expansion for $\lambda_3$ starts at $g^{0}$, for example in free scalar theory with one field \cite{MS}). 

Our findings in this paper are consistent with a picture where  $\lambda_3$ of $\mathcal{N}=4$ SYM plasma starts off non-zero (or zero \cite{Sohrabi}) at vanishing coupling and approaches zero asymptotically as $\lambda^{-3/2}$ with a slope given by our result (\ref{L3R}) at large values of $\lambda$. The possibility of $\lambda_3=0$ at vanishing coupling would be quite interesting, since our analysis would predict a non-monotonic behavior for $\lambda_3$ as a function of the coupling constant $\lambda$. This has to be contrasted with, for instance $\eta/s$ which depends monotonically on the 't Hooft coupling. This remains to be seen and surely calls for further investigations. It would also be nice to find the non-linear second-order coefficients for rotating fluids in AdS/CFT along the lines of \cite{Rotating}.
\section*{Acknowledgment} 
After completing our work, we became aware of related work by Peter Arnold, Diana Vaman, Chaolun Wu and Wei Xiao. We would like to thank Guy D. Moore for valuable discussions. O.S. is grateful to Kevin Schaeffer, Kostas Skenderis and Balt Van Rees. O.S. is supported by the Berkeley Center for Theoretical Physics, department of physics at UC Berkeley and in part by DOE, under contract DE-AC02-05CH11231. K.~S. work is supported in part by the Natural Sciences and Engineering Research Council of Canada.

\end{document}